# On the photosynthetic potential in the very Early Archean oceans

[1] Daile Avila, [2] Rolando Cardenas, [2] Osmel Martin

[1] *Industrial Fishing Enterprise, Ciego de Avila, Cuba*

[2] *Planetary Science Group, Physics Department, Universidad Central de Las Villas, Santa Clara, Cuba*

Phone 53 42 281109

Fax 53 42 281130

e-mail: rcardenas@uclv.edu.cu

**Abstract**: In this work we apply a mathematical model of photosynthesis to quantify the potential for photosynthetic life in the very Early Archean oceans. We assume the presence of oceanic blockers of ultraviolet radiation, specifically ferrous ions. For this scenario, our results suggest a potential for photosynthetic life greater than or similar to that in later eras/eons, such as the Late Archean and the current Phanerozoic eon.

# 1. Introduction

Seemingly, the origin of photosynthetic life took place approximately 3.5 Ga ago, during the Archean Eon. There are three main lines of evidence for the antiquity of photosynthesis: chemical markers, microfossils of ancient organisms and stromatolite fossils (Olson and Blankenship 2004). Among the chemical markers the ratio of $^{13}C/^{12}C$ in sedimentary organic carbon (kerogen) indicates a continuous record of biological $CO_2$ fixation that goes back 3.5–3.8 Ga (Schidlowski et al. 1983; Schidlowski 1988). However, the most persuasive evidence is the existence of the Early Archean stromatolites: layered structures consisting of alternate layers of mat-forming organisms and sediment, extant stromatolites almost always containing filamentous photosynthetic bacteria and/or cyanobacteria. There is also a continuous fossil record of stromatolites from structures dated at 3.1 and 3.5 Ga (Walter 1983), suggesting that phototrophs had already emerged 3500 million years ago (Awramik 1992).

It is stated that the very first organisms capable of doing photosynthesis were anoxygenic, but then were overwhelmed by more efficient photoautotrophs: cyanobacteria. At this time, prior to the formation of the ozone layer and perhaps with no other atmospheric UV blockers, the flux of ultraviolet radiation (UVR) on the ocean surface was higher than today (Singh et al. 2008).

UV radiation has a diversity of damaging effects in microorganisms, from DNA damage to photosynthesis inhibition. Thus, it is often stated that the photobiological regime in the Archean eon (with ozone-less atmosphere) would put a very strong constraint on surface aquatic life. However, an absence of atmospheric UV shields does not necessarily imply an absence of oceanic ones. In the seawater existed several organic and inorganic chemical species that have been proposed as potential UV absorbers in the Early Archean oceans: organic polymers from electric discharges and HCN polymerizations, solubilised elemental sulphur, inorganic ions such as $Cl^-$, $Br^-$, $Mg^{2+}$, and $SH^-$ or $Fe^{2+}$. An oil slick (formed by polymerisation of atmospheric methane under the action of solar UVR) or large quantities of organic foam would also serve as a very efficient UV protection, see Cleaves and Miller (1998) and references therein. The most important inorganic UV absorbers appear to be $Fe^{2+}$ and $H_2S$, with Archean banded iron formations providing some evidence for concentrations of $Fe^{2+}$ much higher than those of $H_2S$. The first has significant absorption in the UV region and was present in the anoxic very Early Archean oceans (Lowe 1982). It would have originated from deep ocean hydrothermal upwelling and could have provided significant UV attenuation for the benthos and mixed layer and intertidal habitats (Olson and Pierson 1986; Garcia-Pichel 1998). However, after the advent of oxygenic photosynthesis, oxygen would have stripped the mixed layer of soluble reduced iron (Drever 1974). Thus, this UV protection would not have existed in what we here call the Late Archean (ocean oxygenated). In Cockell (2000) the possibilities for life under the photobiological regime expected at above mentioned era were presented. In Martin et al. (2011) some of us preliminarily quantified the potential for photosynthesis in an analogous context. In this work we go further back in time to quantify the potential for photosynthesis in the very Early Archean. This scenario, which differs from that in Cockell (2000) and in Martin et al. (2011), assumes a non-oxygenated upper ocean, thus allowing for the presence of $Fe^{2+}$ ions, taking advantage of works recently done in Lake Matano, a modern analogue of the very Early Archean ocean, currently under investigation (Crowe et al. 2008).

## 2. Materials and methods

### 2.1 The solar radiation and atmospheric radiative transfer models

To account for solar spectral irradiances at ground level, both during Early and Late Archean, we used considerations similar to those in Cockell (2000). The more important assumptions are a 25% less luminous Sun (for all wavelengths) and the absence of UV blockers (ozone, sulphur, organic haze) in the atmosphere. Therefore, we use spectral irradiances at ground level as in the above mentioned reference.

To account for the same in the (current) Phanerozoic eon, we used the computer code Tropospheric Ultraviolet and Visible (TUV) Radiation Model (free for download from http://cprm.acd.ucar.edu/Models/TUV/). Solar spectral irradiances at ground level were calculated for a modern Phanerozoic atmosphere with an ozone content of 350 Dobson units.

In all cases (Archean and Phanerozoic) cloudless skies were assumed, and spectral irradiances were obtained for solar zenith angles of 0 and 60 degrees.

## 2.2 The oceanic radiative transfer models

For the Early Archean ocean we considered the presence of $Fe^{2+}$ ions, in concentrations as suggested by Crowe et al. (2008). This implies very strong UV absorption of UV-C and UV-B bands. The attenuation coefficients in the 210 – 310 nm range for an Early Archean ocean model, with $Fe^{2+}$ concentration of $10^{-4}$ mol/dm$^3$, were inferred from data in Cleaves and Miller (1998). The remaining attenuation coefficients (310 – 700 nm) were Smith and Baker (1981) coefficients for clearest waters, as in Cockell (2000). The Late Archean ocean was already stripped from $Fe^{2+}$ ions, thus for all wavelengths (222 – 700 nm) we used Smith and Baker (1981) coefficients. For the current Phanerozoic ocean, we took the attenuation coefficients as in Peñate et al. (2010), selecting water type I, the clearest type of Jerlov's optical ocean water classification (Shifrin 1988).

The spectral irradiances just below sea surface ($z = 0^-$) were obtained from the corresponding ones just above sea surface ($z = 0^+$) through:

$$E(\lambda, 0^-) = [1 - R]E(\lambda, 0^+) \qquad (1)$$

where R is the reflection coefficient obtained from Fresnel formulae applied to the interface air-water. The spectral irradiances $E(\lambda, z)$ at depth $z$ were calculated using Lambert-Beer's law of optics:

$$E(\lambda, z) = E(\lambda, 0^-) \exp[-K(\lambda)z] \qquad (2)$$

$K(\lambda)$ stands for the attenuation coefficient for wavelength $\lambda$.

The irradiances of photosynthetically active radiation (PAR) were calculated by:

$$E_{PAR}(z) = \sum_{\lambda=400nm}^{700nm} E(\lambda, z)\Delta\lambda \qquad (3)$$

while those for ultraviolet radiation (UVR) were weighted with a biological weighting function $\varepsilon(\lambda)$ which gives more weight to more inhibitory wavelengths:

$$E_{UV}^*(z) = \sum_{\lambda=222nm}^{399nm} \varepsilon(\lambda) E(\lambda, z)\Delta\lambda \qquad (4)$$

In both cases above $\Delta\lambda = 1nm$. The biological weighting function $\varepsilon(\lambda)$ used is a generalized action spectrum for temperate phytoplankton (Neale 2011) which accounts for both DNA damage and photosystem inhibition, resulting in whole-cell phytoplankton photosynthesis inhibition. In Fig. 1 we present this action spectrum, normalized respect to its value at $\lambda = 300nm$. Its full wavelength range is from $\lambda = 222nm$ to $\lambda = 399nm$ (even without atmospheric ozone, spectral irradiances below $\lambda = 222nm$ would be very tiny at sea level).

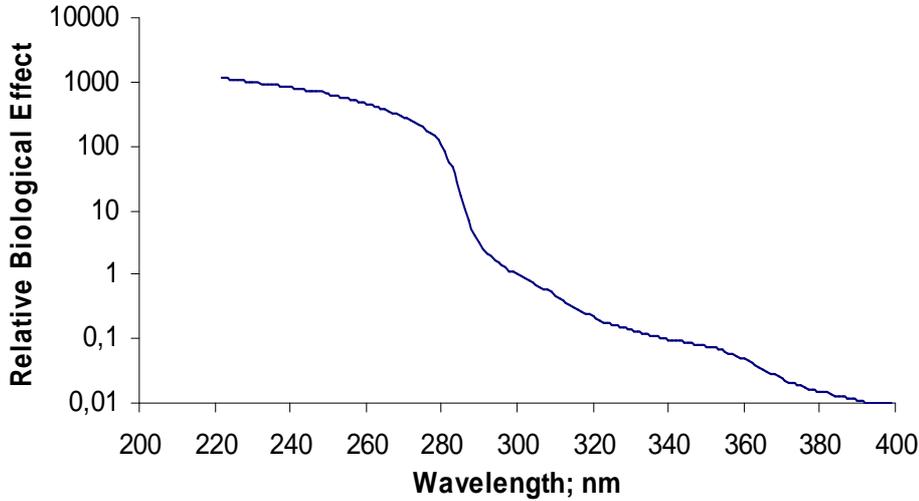

**Fig. 1** The action spectrum used in this paper to evaluate whole-cell phytoplankton photosynthesis inhibition, normalised to its value at $\lambda = 300nm$.

**2.3 The oceanic circulation pattern**

In this first modelling, we chose two circulation patterns: ocean at rest (no motion) and vertical circular currents (Langmuir circulation). Higher Archean temperatures might have implied a shallower upper mixed layer (UML), leading to a greater exposition of phytoplankton to solar UVR. As it is difficult to place constraints to the depth of the mixed layer during the Archean (Cockell 2000), we assume Langmuir cells of different depths: 20, 40 and 100 meters, and make a comparison of average photosynthesis rates.

**2.4 The photosynthesis model**

To compute the photosynthesis rates $P$ at depth $z$ (normalised to saturation rates $P_S$), we use a model for photosynthesis, typically employed with phytoplankton assemblages with good repair capabilities (Fritz et al. 2008):

$$\frac{P}{P_S}(z) = \frac{1 - \exp[-E_{PAR}(z)/E_S]}{1 + E_{UV}^*(z)} \qquad (5)$$

where $E_{PAR}(z)$ and $E_{UV}^*(z)$ are the irradiances of photosynthetically active radiation (PAR) and ultraviolet radiation (UVR) at depth $z$. The parameter $E_S$ accounts for the efficiency with which the species uses PAR: the smaller its value, the greater the efficiency using PAR. The asterisk in $E_{UV}^*(z)$ means that spectral irradiances of UVR are weighted with a biological weighting function $\varepsilon(\lambda)$.

The average photosynthesis rate $\langle P/P_S \rangle$ in the upper mixed layer was calculated splitting this layer into $N$ thinner ones. Then, the photosynthesis rate $P/P_S(n)$ at the middle depth inside the n-th layer was calculated. The average photosynthesis rate in the UML is given by:

$$\left\langle \frac{P}{P_S} \right\rangle = \frac{\sum_{n=1}^{N} \frac{P}{P_S}(n)}{N} \qquad (6)$$

## 3. Results

Figures 2-7 show the photosynthesis rates in representative photic zones of Early Archean, Late Archean and Phanerozoic eon. As expected, our plots show that UVR is an important environmental stressor only in the first tens of meters and species more efficiently using PAR (those with smaller value of the parameter $E_S$), will always show higher photosynthesis rates, as also shown by some of us in Martin et al. (2011). Therefore, in most contexts analysed in this paper, we see that photosynthesis starts with low rates near the water surface (too much UV), then increases down the water column (as the UVR stress decreases) until a maximum is reached. The only exception to the above mentioned general trend is seen in Figure 7, which shows the photobiological regime in the Phanerozoic eon for a solar zenith angle of 60 degrees. In this case, low and intermediate efficient species ($E_S$ = 100 W/m$^2$ and $E_S$ = 50 W/m$^2$) start with their maximum photosynthesis rate at the surface.

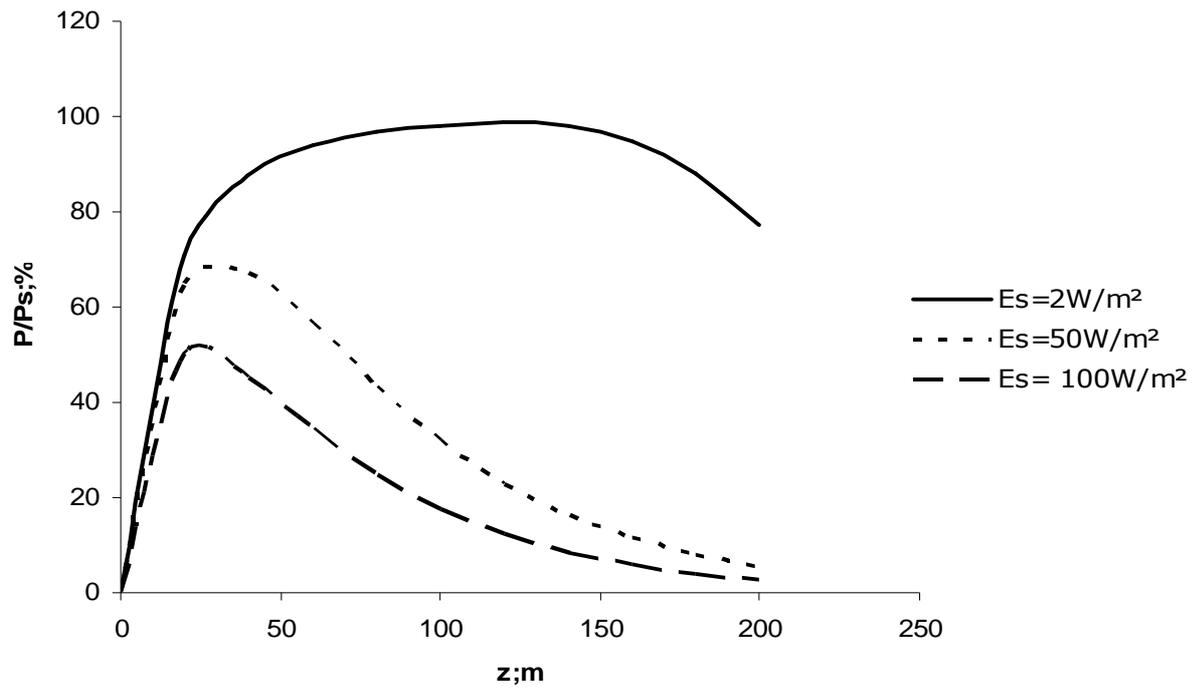

**Fig. 2** Photosynthesis rates in Early Archean for a solar zenith angle of 0° considering the photoprotection of $Fe^{2+}$.

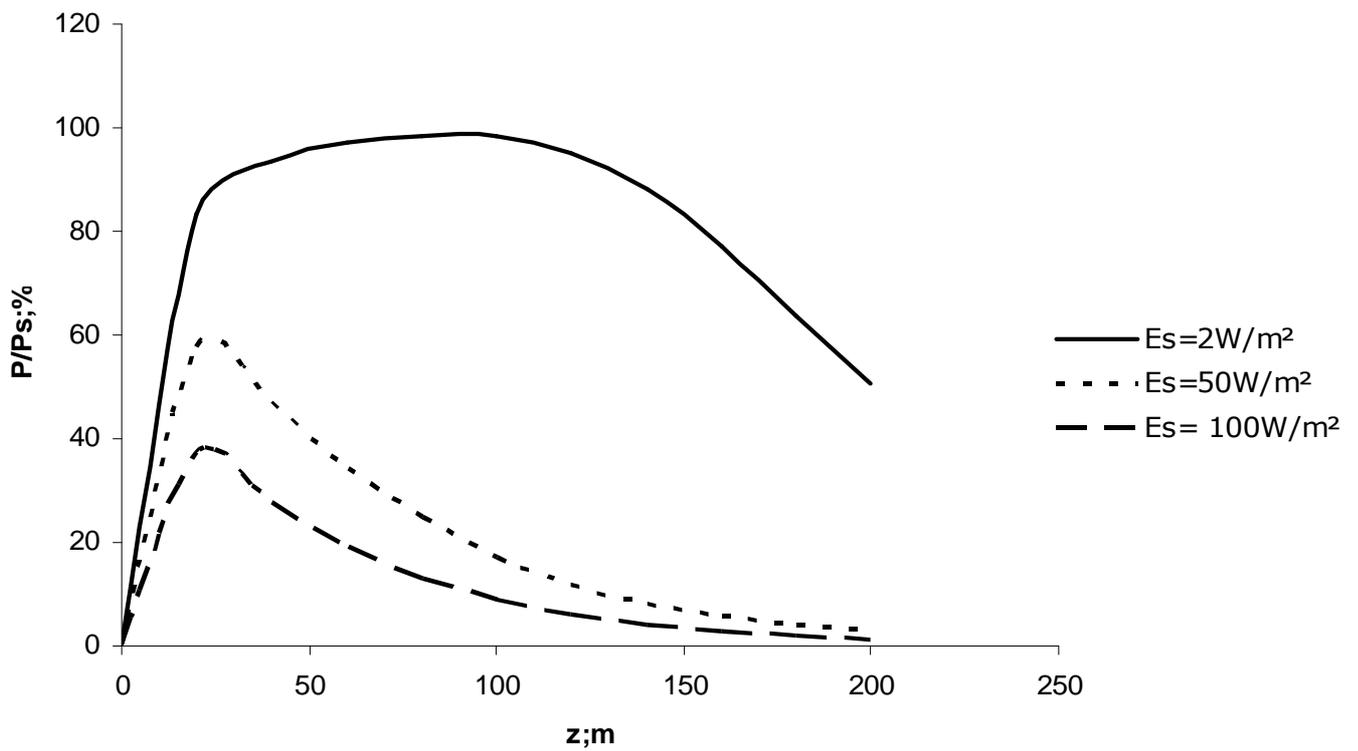

**Fig. 3** Photosynthesis rates in Early Archean for a solar zenith angle of 60° considering the photoprotection of $Fe^{2+}$.

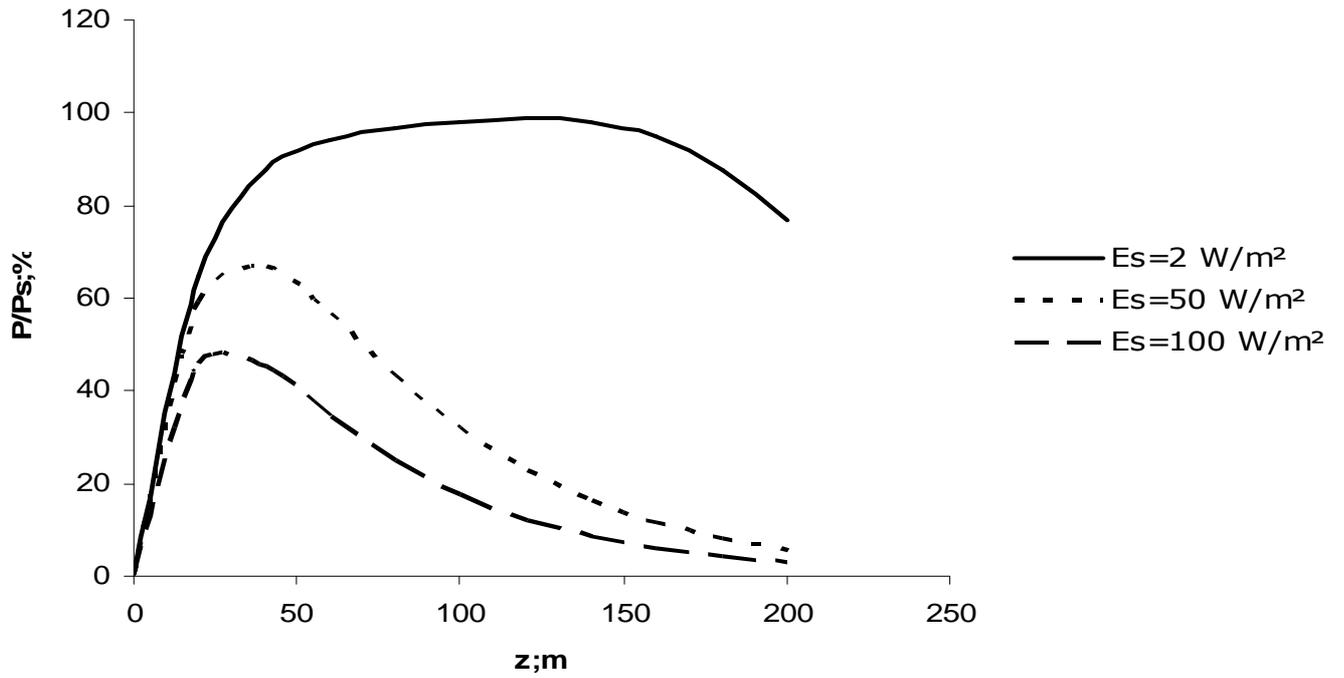

**Fig. 4** Photosynthesis rates in Late Archean for a solar zenith angle of 0°.

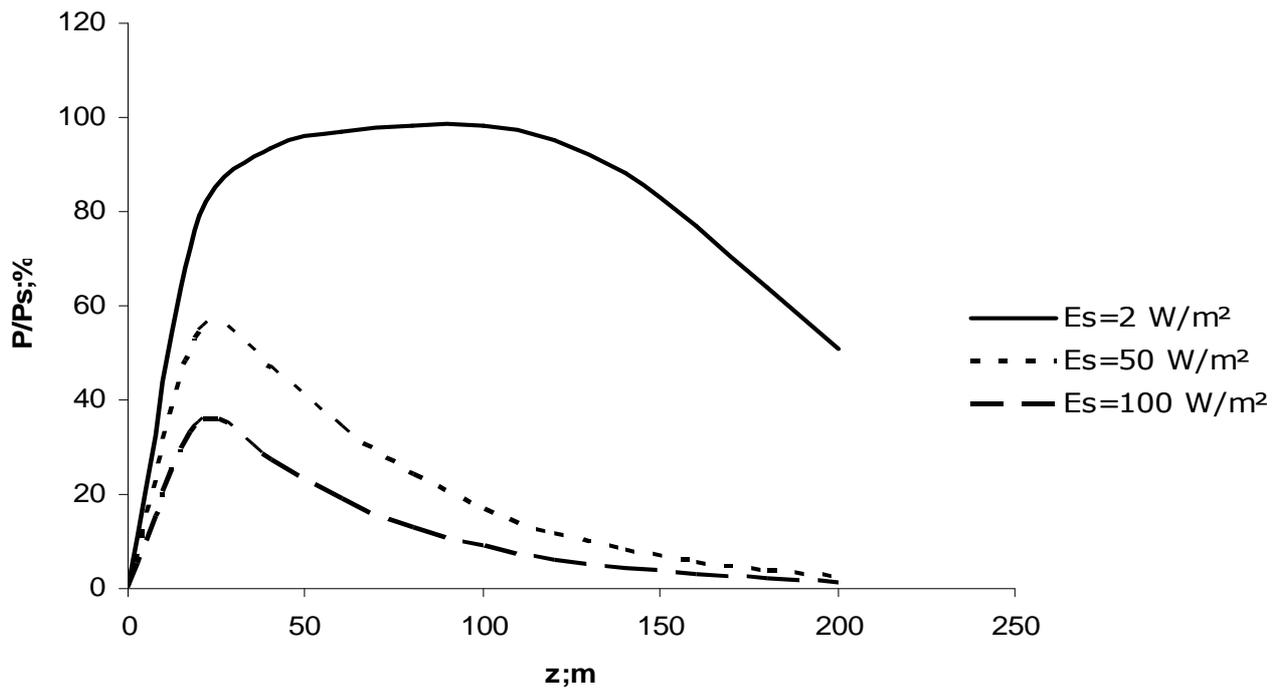

**Fig. 5** Photosynthesis rates in Late Archean for a solar zenith angle of 60°.

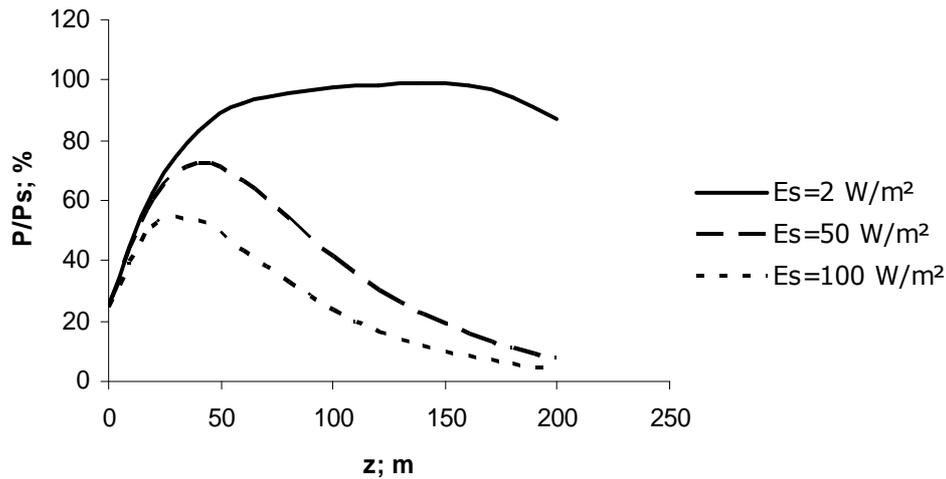

**Fig. 6** Photosynthesis rates during Phanerozoic eon for a solar zenith angle of 0° and ocean optical water type I

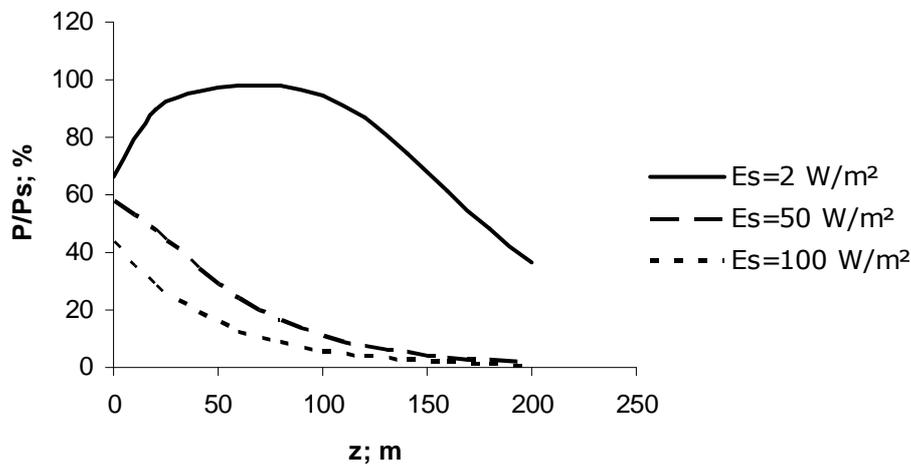

**Fig. 7** Photosynthesis rates during Phanerozoic eon for a solar zenith angle of 60° and ocean optical water type I

As phytoplankton can easily be trapped by currents in the upper mixed layer, it is exposed to different levels of UVR and PAR, and then it is instructive to quantify the average photosynthesis rate in a given UML. Using the procedure explained in subsection 2.4, we calculated the average photosynthesis rates for three Langmuir cells having mixing depths of 20, 40 and 100meters. In Table 1 we present the results. Comparing the Early Archean with the other two scenarios, in all cases the photosynthesis rates are higher than in the Late Archean, and in several cases higher than during the Phanerozoic.

Table 1 Average photosynthesis rates in three different Langmuir cells

| Depth | | Es=2 W/m² | | Es=50 W/m² | | Es=100 W/m² | |
|---|---|---|---|---|---|---|---|
| | | 0° | 60° | 0° | 60° | 0° | 60° |
| 20 m | Early Archean | 53.9 | 70.4 | 51.8 | 56.6 | 43.8 | 39.9 |
| | Late Archean | 24.5 | 35.2 | 23.1 | 26.6 | 18.7 | 17.9 |
| | Phanerozoic | 50.3 | 89.6 | 49.8 | 45.3 | 45.7 | 28.0 |
| 40 m | Early Archean | 67.2 | 80.0 | 59.8 | 54.8 | 46.3 | 36.2 |
| | Late Archean | 51.8 | 61.9 | 44.8 | 39.5 | 33.3 | 25.0 |
| | Phanerozoic | 62.3 | 95.6 | 59.2 | 26.1 | 49.8 | 15.8 |
| 100 m | Early Archean | 88.2 | 88.5 | 56.6 | 52.1 | 37.2 | 31.4 |
| | Late Archean | 81.7 | 86.0 | 50.3 | 35.1 | 33.0 | 20.7 |
| | Phanerozoic | 85.0 | 96.7 | 62.0 | 37.4 | 45.2 | 26.3 |

## 4. Discussion

Above plots show that in general UVR is an important stressor in the first tens of meters of the water column. This is due to the stronger attenuation, compared to PAR, which ocean water exerts for UVR (Peñate et al. 2010). At a given depth, which depends on the optical water type, solar zenith angle and efficiency of the species to use PAR, the optimum balance of the actions of UVR and PAR is achieved. These are the peaks or maxima in Figures 2-7. Deeper in the water column the effect of the scarcity of PAR is more important, and thus photosynthesis rates decrease. In general these maxima are smaller and are reached at lower depths for low efficient species, as for them the scarcity of PAR (which they do not use very well) is more important than the UVR stress.

From Table 1 we check that there is not an absolute general trend for all the three scenarios because there is the confluence of several variables and parameters: intensities of UVR and PAR, the efficiency in the use of PAR (given by $E_S$) and the solar zenith angle. However, in all cases we see a better photosynthetic potential in Early Archean compared to Late Archean. This is to be expected, as the only environmental difference is the presence of the protective $Fe^{2+}$ ions in the former and its absence in the latter. Comparison of the Early Archean with the Phanerozoic is a more complicated, but at first glance the possibilities in the Early Archean do not look inferior, if a similar depth of the UML is assumed. However, it is generally accepted that the average temperature during the Archean was higher than during the Phanerozoic. How higher is still a matter of debate, so it is difficult to estimate the average upper mixed layer depth during the Archean. In general, a shallower UML is to be expected during the Archean, implying greater exposure to UVR, which tends to lower photosynthesis rates. Table 1 gives us an idea on the dependence of the photosynthesis rates with UML depth.

## 5. Conclusions

Our main conclusion is that the presence of an oceanic UV blocker such as $Fe^{2+}$ would have greatly ameliorated the photobiological regime in the Early Archean, creating an environment with a potential for photosynthesis definitely better than that during the Late Archean. In general, photosynthesis rates would have average values in the upper layer of the ocean (where UVR is a stressor) higher than in the Late Archean. The presence of other oceanic UV blockers would still improve this situation, provided their PAR absorption was negligible.

Comparison with the current Phanerozoic eon gets complicated, because of the higher average temperatures during the Archean, which might have implied shallower mixed layer depths and greater exposure to UVR. If similar vertical mixing patterns are assumed, then photosynthetic potential during Early Archean and Phanerozoic eon do not look very different.